\documentclass[prd,nofootinbib,showpacs,preprint]{revtex4}
\usepackage{amsmath}
\usepackage{graphicx}
\usepackage{epsfig}
\graphicspath{{Figs/}}
\usepackage{dcolumn}
\usepackage{bm}
\usepackage{amssymb}
\usepackage[usenames,dvipsnames]{color}
\usepackage{slashed}
\usepackage[dvipdfm,colorlinks,citecolor=blue]{hyperref}

\begin{document}
\title{Spontaneous Parity Breaking, Gauge Coupling Unification and Consistent Cosmology with Transitory Domain Walls}
\author{Debasish Borah}
\email{dborah@tezu.ernet.in}
\affiliation{Department of Physics, Tezpur University, Tezpur-784028, India}

\begin{abstract}
Formation of transitory domain walls is quite generic in theories with spontaneous breaking of discrete symmetries. Since these walls are in conflict with cosmology, there has to be some mechanism which makes them disappear. We study one such mechanism by incorporating Planck scale suppressed operators within the framework of Left-Right Symmetric Models (LRSM) where Left-Right parity (D parity) is spontaneously broken. We find that this mechanism can not make the walls disappear in minimal versions of LRSM. We propose two viable extensions of this model and show that Planck scale suppressed operators can give rise to successful disappearance of domain walls provided the scale of parity breaking obeys certain limits. We also constrain the scale of parity breaking by demanding successful gauge coupling unification and make a comparison of the unification and cosmology bounds.
\end{abstract}

\maketitle

\section{Introduction}
Spontaneous parity breaking can be naturally explained within the framework of Left-Right Symmetric Models (LRSM) \cite{lrsm} which have been considered a novel extension of Standard Model (SM) and studied extensively for last few decades. Also, tiny neutrino masses \cite{PDG} can be successfully implemented via seesaw mechanism \cite{ti, tii} without reference to very high scale physics such as grand unification. Incorporating supersymmetry into such models have several other motivations like protecting the scalar sector from quadratic divergences, providing a natural dark matter candidate among others. However, as studied previously in \cite{Mishra:2009mk, borah}, generic Supersymmetric Left-Right models are tightly constrained from consistent cosmology as well as successful gauge coupling unification point of view and in quite a few cases these models do not give rise to successful unification and consistent cosmology simultaneously. With a view of this, we intend to study non-supersymmetric versions of LRSM and check if both of these constraints can be satisfied simultaneously.

Spontaneous breaking of exact discrete symmetries like parity (which we shall denote as
D-parity hereafter) has got cosmological
implications since they lead to frustrated phase transitions leaving behind a network of domain walls (DW). These 
domain walls, if not removed will be in conflict with the observed Universe \cite{Kibble:1980mv,Hindmarsh:1994re}. It was pointed 
out \cite{Rai:1992xw,Lew:1993yt} that Planck scale suppressed non-renormalizable operators 
can be a source of domain wall instability. Gauge structure of the underlying theory dictates the structure of these non-renormalizable operators. In presence of supersymmetry, the constraints on the D-parity breaking scale was discussed in \cite{Mishra:2009mk}. Here we perform the same analysis in the absence of supersymmetry and also check if successful gauge coupling unification can be achieved simultaneously in these models. Earlier studies on gauge coupling unification in non-supersymmetric LRSM were done in \cite{LRunif, Lindner} and very recently in \cite{siringo}. Here adopt the procedure outlined in \cite{Lindner} for gauge coupling unification and that in \cite{Mishra:2009mk} for domain wall disappearance mechanism to show that in the minimal versions of non-supersymmetric LRSM, successful gauge coupling unification and domain wall disappearance can not be achieved simultaneously. We then propose two possible extensions of such minimal models which can successfully give rise to both of these desired outcomes.

It is worth mentioning that the formation of domain walls is 
not generic in all Left-Right models. Models where D-parity and $SU(2)_R$ gauge symmetry are broken at two different stages do not suffer from this problem \cite{Chang}. In these models, the vacuum expectation value (vev) of a parity odd singlet field breaks the D-parity first and $SU(2)_R$ gauge symmetry gets broken at a later stage by either Higgs triplets or Higgs doublets.

This paper is organized as follows. In section \ref{sec:DW-section} we briefly review the 
domain wall dynamics. In section \ref{minLR} we discuss minimal LRSM with Higgs doublets 
 and with Higgs triplets,and discuss how the requirement of successful disappearance of domain walls gives rise to unphysical constraints on the parity breaking scale. In section \ref{extLR} we discuss two viable extensions of the minimal LRSM which can provide a viable solution to the domain wall problem. Then in section \ref{GUT}, we study gauge coupling unification in both minimal and extended LRSM and finally summarize our results in section \ref{conclude}.

\section{Domain Wall dynamics}
\label{sec:DW-section}

Discrete symmetries and their spontaneous breaking are both common instances and desirable in particle physics model building. The spontaneous breaking of such discrete symmetries gives rise to a network of 
domain walls leaving the accompanying phase transition frustrated \cite{Kibble:1980mv, Hindmarsh:1994re}. 
The danger of a frustrated phase transition can therefore be evaded if a small explicit breaking of discrete 
symmetry can be introduced.

If the amount of such discrete symmetry breaking is small, the resulting domain walls may be relatively long lived and
 can dominate the Universe for a sufficiently long time. However, this will be in conflict with the observed Universe and hence these domain 
walls need to disappear at a very high energy scale (at least before Big Bang Nucleosynthesis). In view of this, we summarize the three cases 
of domain wall dynamics discussed in \cite{Mishra:2009mk}, one of which originates in radiation dominated (RD) Universe and destabilized also within
the radiation dominated Universe. This scenario was originally proposed by Kibble \cite{Kibble:1980mv} and Vilenkin \cite{Vilenkin:1984ib}. The second scenario was
essentially proposed in \cite{Kawasaki:2004rx}, which consists of the walls
originating  in a radiation dominated phase, subsequent to which the 
Universe enters a matter dominated (MD) phase, either due to substantial 
production of heavy unwanted relics such as moduli, or simply due
to a coherent oscillating scalar field. The third one is a variant of the 
MD model in which the domain walls dominate the Universe for a considerable epoch giving rise to a mild inflationary behavior or weak inflation (WI) \cite{Lyth:1995hj,Lyth:1995ka}. In all these cases the domain
walls disappear before they come to dominate the energy density of the Universe.

When a scalar field $\phi$ acquires a vev at a scale $M_R$ at some critical temperature $T_c$, a phase transition occurs leading to the formation of domain walls. The energy density trapped per unit area of such a wall is $\sigma \sim  M^3_R$. The dynamics of the walls are determined by two quantities, force due to tension $f_T \sim \sigma/R$ and force due to friction $f_F \sim \beta T^4$ where $R$ is the average scale of radius of curvature prevailing in the wall complex, $\beta$ is the speed at which the domain wall is navigating through the medium and $T$ is the temperature. The epoch at which these two forces balance each other sets the time scale $t_R \sim R/\beta $. Putting all these together leads to the scaling law for the growth of the scale $R(t)$:
\begin{equation}
R(t) \approx (G \sigma)^{1/2}t^{3/2}
\end{equation}
The energy density of the domain walls goes as $\rho_W \sim (\sigma R^2/R^3) \sim (\sigma/Gt^3)^{1/2} $. In a radiation dominated era this $\rho_W$ is comparable to the energy density of the Universe $[\rho \sim 1/(Gt^2)]$ around time $t_0 \sim 1/(G \sigma)$. 

The pressure difference arising from small asymmetry on the two sides of the wall competes with the two forces $f_F \sim 1/(Gt^2)$ and $f_T \sim (\sigma/(Gt^3))^{1/2}$ discussed above. For $\delta \rho$ to exceed either of these two quantities before $t_0 \sim 1/(G\sigma)$
\begin{equation}
 \delta \rho \geq G\sigma^2 \approx \frac{M^6_R}{M^2_{Pl}} \sim M_R^4 \left(\frac{M_R}{M_{Pl}}\right)^2
\label{eq:RD-delta-rho}
\end{equation}

Similar analysis in the matter dominated era, originally considered in \cite{Kawasaki:2004rx} begins with the assumption that the initially formed wall complex in a phase transition is expected to rapidly relax to a few walls per horizon volume at an epoch characterized by Hubble parameter value $H_i$. Thus the initial energy density of the wall complex is $\rho^{\text{in}}_W \sim \sigma H_i$. This epoch onward the energy density of the Universe is assumed to be dominated by heavy relics or an oscillating modulus field and in both the cases the scale factor grows as $a(t) \propto t^{2/3}$. The energy density scales as $\rho_{\text{mod}} \sim \rho^{\text{in}}_{\text{mod}}/(a(t))^3$. If the domain wall (DW) complex remains frustrated, i.e. its energy density contribution $\rho_{\text{DW}} \propto 1/a(t)$, the Hubble parameter at the epoch of equality of DW contribution with that of the rest of the matter is given by \cite{Kawasaki:2004rx}
\begin{equation}
H_{\text{eq}} \sim \sigma^{3/4}H^{1/4}_i M^{-3/2}_{Pl}
\label{heq}
\end{equation}
Assuming that the domain walls start decaying as soon as they dominate the energy density of the Universe, which corresponds to a temperature $T_D$ such that $H^2_{\text{eq}} \sim GT^4_D$, the above equation gives 
\begin{equation}
T^4_D \sim \sigma^{3/2} H^{1/2}_i M^{-1}_{Pl}
\label{td}
\end{equation}
Under the assumption that the domain walls are formed at $T \sim \sigma^{1/3}$
\begin{equation}
H^2_i = \frac{8\pi}{3}G\sigma^{4/3} \sim \frac{\sigma^{4/3}}{M^2_{Pl}}
\end{equation}
Now from Eq. (\ref{td})
\begin{equation}
T^4_D \sim \frac{\sigma^{11/6}}{M^{3/2}_{Pl}} \sim \frac{M^{11/2}_R}{M^{3/2}_{Pl}} \sim M^4_R \left(\frac{M_R}{M_{Pl}}\right)^{3/2}
\end{equation}
Demanding $\delta \rho > T^4_D$ leads to
\begin{equation}
\delta \rho >  M^4_R \left(\frac{M_R}{M_{Pl}}\right)^{3/2}
\end{equation}

The third possibility is the walls dominating the energy density of the Universe for a limited epoch which leads to a mild inflation. This possibility was considered in \cite{Lyth:1995hj,Lyth:1995ka}. As discussed in \cite{Mishra:2009mk}, the evolution of energy density of such walls can be expressed as 
\begin{equation}
\rho_{\text{DW}}(t_d) \sim \rho_{\text{DW}}(t_{\text{eq}}) (\frac{a_{\text{eq}}}{a_d}) 
\end{equation}
where $a_{\text{eq}} (a_d)$ is the scale factor at which domain walls start dominating (decaying) and $t_{\text{eq}} (t_d)$ is the corresponding time. If the epoch of domain wall decay is characterized by temperature $T_D$, then $\rho_{\text{DW}} \sim T^4_D$ and the above equation gives 
\begin{equation}
T^4_D = \rho_{\text{DW}} (t_{\text{eq}}) (\frac{a_{\text{eq}}}{a_d}) 
\label{td1}
\end{equation}
In the matter dominated era the energy density of the moduli fields scale as 
\begin{equation}
\rho^d_{\text{mod}} \sim \rho^{\text{eq}}_{\text{mod}} (\frac{a_{\text{eq}}}{a_d})^3
\end{equation}
Using this in equation (\ref{td1}) gives 
\begin{equation}
\rho^d_{\text{mod}} \sim \frac{T^{12}_D}{\rho^2_{\text{DW}} (t_{\text{eq}})}
\end{equation}
Domain walls start dominating the Universe after the time of equality,  $\rho_{\text{DW}}(t_d) > \rho^d_{\text{mod}} $. So the pressure difference across the walls when they start decaying is given by 
\begin{equation}
\delta \rho \geq \frac{T^{12}_D G^2}{H^4_{\text{eq}}}
\end{equation}
where $H^2_{\text{eq}} \sim G \rho_{\text{DW}} (t_{\text{eq}})$. Replacing the value of $H_{\text{eq}}$ from equation (\ref{heq}), the pressure difference becomes 
\begin{equation}
\delta \rho \geq M^4_R \frac{T^{12}_D M^3_{Pl}}{M^{15}_R}
\end{equation}
Unlike the previous two cases RD and MD, here it will not be possible to estimate $T_D$ in terms of other mass scales and we will keep it as undetermined.

\section{Minimal Left Right Symmetric Model}
\label{minLR}

\subsection{LRSM with Higgs doublets}
We first study left-right symmetric extension of the standard
model with only Higgs doublets. In addition to the usual fermions of the
standard model, we require the right-handed neutrinos to complete the
representations. One of the important features of the model is that it
allows spontaneous parity violation. The Higgs representations then
require a bidoublet field, which breaks the electroweak symmetry and
gives masses to the fermions. But the neutrinos can have Dirac masses
only, which are then expected to be of the order of charged fermion masses.
To implement the seesaw mechanism and obtain the observed tiny masses
of the standard model neutrinos naturally, one has to introduce fermion triplets to give 
rise to the so called type III seesaw mechanism \cite{tiii}. However, we shall restrict ourselves to the scalar sector and shall not
discuss the implications of such additional fermions.

The particle content of the Left-Right symmetric model with Higgs doublet is
$${\rm Fermions:}~~  Q_L \equiv (3,2,1,1/3),~~  Q_R \equiv (3,1,2,1/3), ~~
\Psi_L \equiv (1,2,1,-1),~~  \Psi_R \equiv (1,1,2,-1) $$
$${\rm Scalars:} \quad \Phi \equiv (1,2,2,0), \quad H_L \equiv (1,2,1,1),
\quad H_R \equiv (1,1,2,1)$$
where the numbers in the brackets are the quantum numbers corresponding to the
gauge group $SU(3)_C \times SU(2)_L \times SU(2)_R \times U(1)_{B-L} $.
In addition to the bi-doublet scalar field $\Phi$, we should also have two
doublet fields $H_L$ and $H_R$ to break the left-right symmetry and contribute
to the neutrino masses. Thus the
symmetry breaking pattern becomes
$$ SU(2)_L \times SU(2)_R \times U(1)_{B-L} \quad \underrightarrow{\langle
H_R \rangle} \quad SU(2)_L\times U(1)_Y \quad \underrightarrow{\langle \Phi \rangle} \quad U(1)_{em}$$

\subsection{LRSM with Higgs triplets}
In this section we briefly outline left-right symmetric models with 
different field contents than the one in the previous section. The usual fermions, including the right-handed
neutrinos, belong to the similar
representations as in the previous section. However, the scalar sector now
contains triplet Higgs scalars in addition to the bidoublet Higgs scalar to
break the left-right symmetry. The triplet Higgs scalars can then give
Majorana masses to the standard model neutrinos by the so called type II seesaw 
mechanism \cite{tii}. 

The particle content of LRSM with Higgs triplets is
$${\rm Fermions:}~~  Q_L \equiv (3,2,1,1/3),~~  Q_R \equiv (3,1,2,1/3), ~~
\Psi_L \equiv (1,2,1,-1),~~  \Psi_R \equiv (1,1,2,-1) $$
$${\rm Scalars:} \quad
\Phi \equiv (1,2,2,0), \quad \Delta_L \equiv (1,3,1,2), \quad \Delta_R \equiv (1,1,3,2)$$
The symmetry breaking pattern in this model remains the same as in the previous
model although the structure of neutrino masses changes. In the symmetry breaking
pattern, the scalar $\Delta_R$ now replaces the role of $H_R$, but otherwise
there is no change. 

$$ SU(2)_L \times SU(2)_R \times U(1)_{B-L} \quad \underrightarrow{\langle
\Delta_R \rangle} \quad SU(2)_L\times U(1)_Y \quad \underrightarrow{\langle \Phi \rangle} \quad U(1)_{em}$$

\subsection{Constraints on $M_R$ from domain wall disappearance}
In both the versions of minimal LRSM discussed above, the non zero vev of the right handed
doublet (or triplet) field breaks both $SU(2)_R\times U(1)_{B-L}$ gauge symmetry as well as the discrete left-right parity (or D parity) and hence gives rise to transitory domain walls. In this section, we consider explicit D parity breaking Planck suppressed operators in these models by adopting the technique developed in\cite{Mishra:2009mk}. And we find constraints
on the parity breaking scale by demanding that these Planck suppressed operators give rise to successful disappearance of domain walls.

In both the minimal versions of LRSM discussed above, the leading non-renormalizable 
operator is of dimension six which can be written as
\begin{equation}
V_{\text{NR}} \supset f_L \frac{(\Sigma^{\dagger}_L \Sigma_L )^3}{M^2_{Pl}} + f_R \frac{(\Sigma^{\dagger}_R \Sigma_R )^3}{M^2_{Pl}}
\end{equation}
where $\Sigma$ can either be a Higgs doublet or a Higgs triplet. Assuming a phase where only right type fields get non-zero vev and left type
fields get zero vev, the scalar potential up to the leading term in $1/M^2_{Pl}$ becomes
\begin{equation}
V^R_{eff} \sim \frac{f_R}{M^2_{Pl}}M^6_R
\end{equation}
Similarly assuming non-zero vev for left type 
fields only and not for right type fields the effective potential becomes
\begin{equation}
 V^L_{eff} \sim \frac{f_L}{M^2_{Pl}}M^6_L
 \end{equation}
Due to the equal chance of both $\Sigma_L$ and $\Sigma_R$ acquiring the same vev (guaranteed by the presence of discrete left-right symmetry), we consider $M_L = M_R$. Thus, the effective energy difference across the walls separating these two vacua is given by
\begin{equation}
\delta \rho \sim \frac{(f_L-f_R)}{M^2_{Pl}}M^6_R
\end{equation}
Now we shall compare this $\delta \rho$ with the case in a matter dominated era
where we have calculated the energy density for the domain wall to disappear.
\begin{equation}
 \frac{(f_L-f_R)}{M^2_{Pl}}M^6_R > M_R^4 \left( 
 \frac{M_R}{M_{Pl}}\right)^{3/2}
\end{equation}
Taking the dimensionless parameters $f$ to be of order unity, the above equation gives a lower bound
on $M_R$ in a matter dominated era
\begin{equation}
M_R > M_{Pl}
\end{equation}
which is unnatural considering the fact that Planck scale is the maximum energy scale a physical theory can have. 
Similarly for radiation dominated era
\begin{equation}
 \frac{(f_L-f_R)}{M^2_{Pl}}M^6_R >M_R^4 \left( 
 \frac{M_R}{M_{Pl}}\right)^2
\end{equation}
which does not give a bound on $M_R$. Rather it gives a lower bound $f_L - f_R >1$. This is also unnatural since dimensionless
 couplings are generically taken to be of order one.
 
Thus, in both the versions of minimal LRSM discussed above, the requirement of domain wall disappearance gives rise to unnatural constraints 
on the scale of parity breaking and the dimensionless parameters. This drawback of such minimal models appeals for suitable extensions 
so as to guarantee successful disappearance of cosmologically unwanted domain walls. In this work, we propose two such possible extensions as discussed in the next section.

\section{Extended LRSM and Successful Disappearance of Domain Walls}
\label{extLR}
In this section, we propose two viable as well as minimal extensions of the left right symmetric models discussed above. We show that, such extensions
 can give rise to successful disappearance of domain walls and different symmetry breaking patters as well as phenomenology.
\subsection{Extension by a gauge singlet}
Consider a gauge singlet field $\sigma (1,1,1,0) $ which is even under parity so that the tree level Lagrangian is parity symmetric until the $\Sigma_R$ field acquires a vev to break (spontaneously) parity as well as the $SU(2)_R \times U(1)_{B-L}$ gauge symmetry to $U(1)_Y$ of standard model. To study the domain wall disappearance mechanism in this model, we consider Planck suppressed higher dimensional operators as in the previous section. Unlike before, here we can have dimension five Planck suppressed terms in the scalar potential which, as we will see, create sufficient pressure difference across the domain walls to make them disappear. These operators can be written as
\begin{equation}
V_{\text{NR}} \supset f_L \sigma \frac{(\Sigma^{\dagger}_L \Sigma_L )^2}{M_{Pl}} + f_R \sigma \frac{(\Sigma^{\dagger}_R \Sigma_R )^2}{M_{Pl}}
\end{equation}
where $\Sigma$ can either be a Higgs doublet or a Higgs triplet and $\sigma$ is the gauge singlet we have introduced. Since a singlet like $\sigma (1,1,1,0)$ can naturally fit inside several $SO(10)$ representations, we assume the vev of this singlet field to be of order $\langle \sigma \rangle \sim M_{\text{GUT}} \sim 2\times 10^{16} \; \text{GeV} $. Assuming a phase where only right type fields get non-zero vev and left type
fields get zero vev, the scalar potential up to the leading term in $1/M_{Pl}$ becomes
\begin{equation}
V^R_{eff} \sim \frac{f_R}{M_{Pl}}M_{\text{GUT}}M^4_R
\end{equation}
Similarly assuming non-zero vev for left type 
fields only and not for right type fields the effective potential becomes
\begin{equation}
 V^L_{eff} \sim \frac{f_L}{M_{Pl}}M_{\text{GUT}}M^4_L
 \end{equation}
Due to the equal chance of both $\Sigma_L$ and $\Sigma_R$ acquiring the same vev, we consider $M_L = M_R$. Thus, the effective energy difference across the walls separating these two vacua is given by
\begin{equation}
\delta \rho \sim \frac{(f_L-f_R)}{M_{Pl}}M_{\text{GUT}}M^4_R
\end{equation}
Comparing this $\delta \rho$ with the case in a matter dominated era, we have
\begin{equation}
 \frac{(f_L-f_R)}{M_{Pl}}M_{\text{GUT}}M^4_R > M_R^4 \left(\frac{M_R}{M_{Pl}}\right)^{3/2}
\end{equation}
Taking the dimensionless parameters $f$ to be of order unity, the above equation gives a upper bound
on $M_R$ in a matter dominated era
\begin{equation}
M_R < (M_{Pl}M^2_{\text{GUT}})^{1/3} \sim 10^{17} \; \text{GeV}
\end{equation}
Similarly for radiation dominated era
\begin{equation}
 \frac{(f_L-f_R)}{M_{Pl}}M_{\text{GUT}}M^4_R >M_R^4 \left(\frac{M_R}{M_{Pl}}\right)^2
\end{equation}
which gives a similar upper bound on $M_R$ as
\begin{equation}
M_R < (M_{Pl}M_{\text{GUT}})^{1/2} \sim 10^{17} \; \text{GeV}
\end{equation}
Comparing the obtained $\delta \rho$ with the weak inflation case we have 
\begin{equation}
\frac{(f_L-f_R)}{M_{Pl}}M_{\text{GUT}}M^4_R \geq  M^4_R \frac{T^{12}_D M^3_{Pl}}{M^{15}_R}
\end{equation}
Taking the dimensionless coefficients to be of order one, we arrive at the following bound on $M_R$
\begin{equation}
M_R \geq  1 \times 10^{4} T^{4/5}_D 
\end{equation}
Thus, for $T_D$ of the electroweak scale, $M_R > 4 \times 10^5 \; \text{GeV}$.
\begin{figure}
\centering
\includegraphics{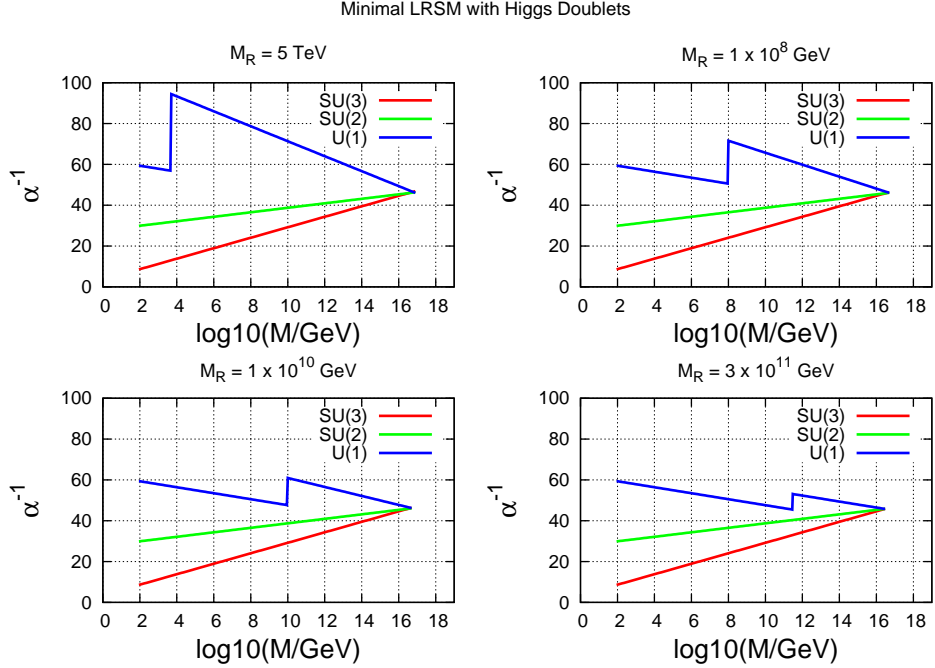}
\caption{Gauge coupling unification in minimal LRSM with Higgs doublets. Including the presence of extra pair of fields $(1,1,1,2), (1,1,1,-2)$ allows the possibility to have low scale $M_R$. The four plots corresponds to number of such extra pairs $n = 6, 4, 2, 0$ respectively}
\label{fig1}
\end{figure}
\begin{figure}
\centering
\includegraphics{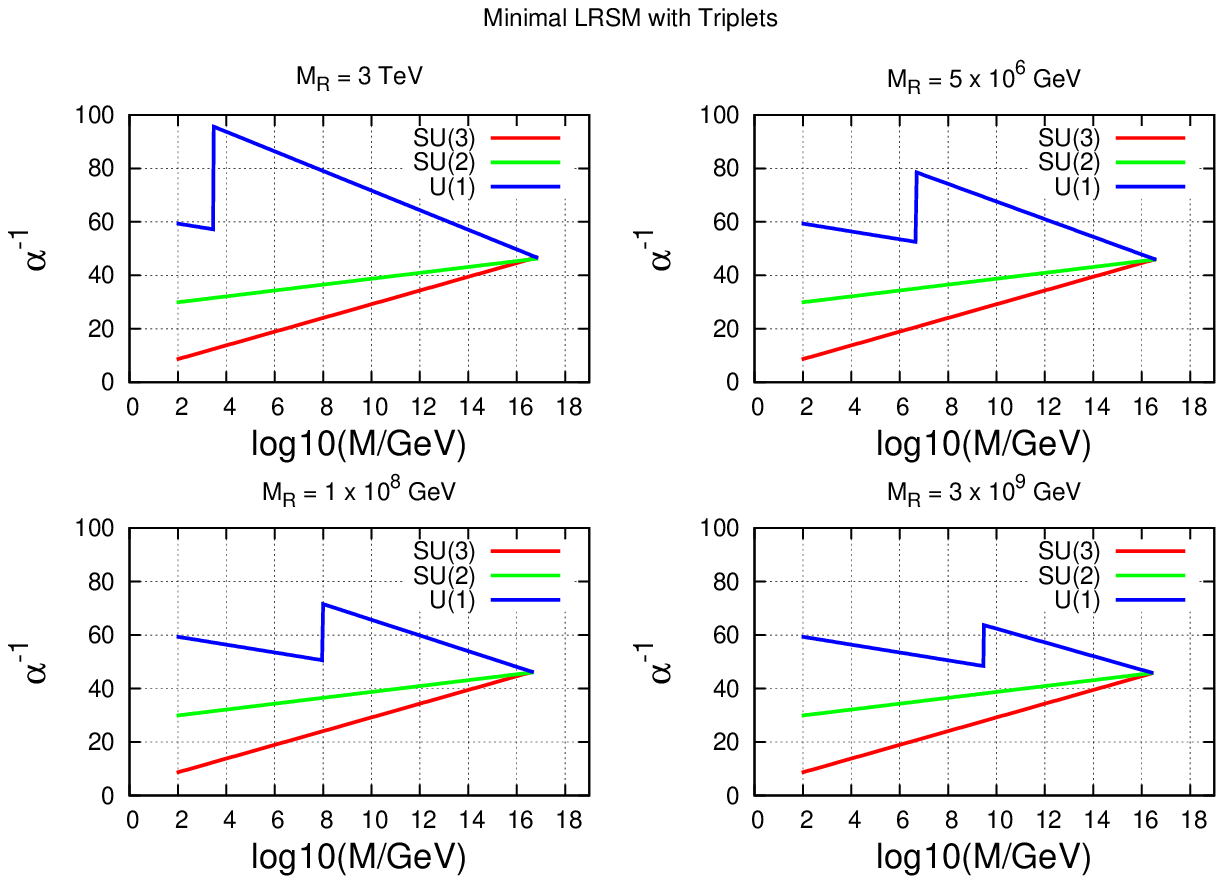}
\caption{Gauge coupling unification in minimal LRSM with Higgs triplets. Including the presence of extra pair of fields $(1,1,1,2), (1,1,1,-2)$ allows the possibility to have low scale $M_R$. The four plots corresponds to number of such extra pairs $n = 3, 2, 1, 0$ respectively}
\label{fig2}
\end{figure}
\begin{figure}
\centering
\includegraphics{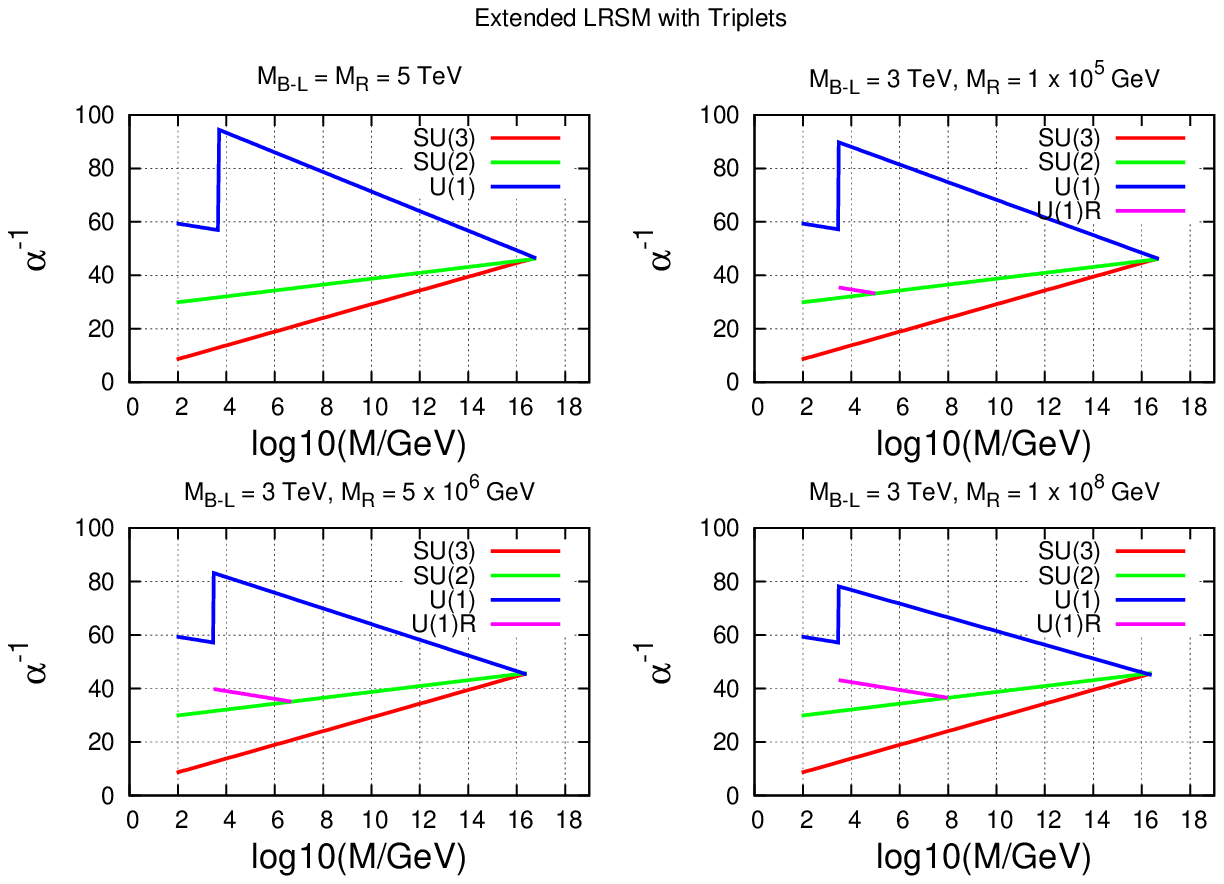}
\caption{Gauge coupling unification in extended LRSM with Higgs triplets. Including the presence of extra pair of fields $(1,1,1,2), (1,1,1,-2)$ allows the possibility to have low scale $M_R$. The four plots corresponds to number of such extra pairs $n = 3, 2, 1, 0$ respectively}
\label{fig3}
\end{figure}
\begin{figure}
\centering
\includegraphics{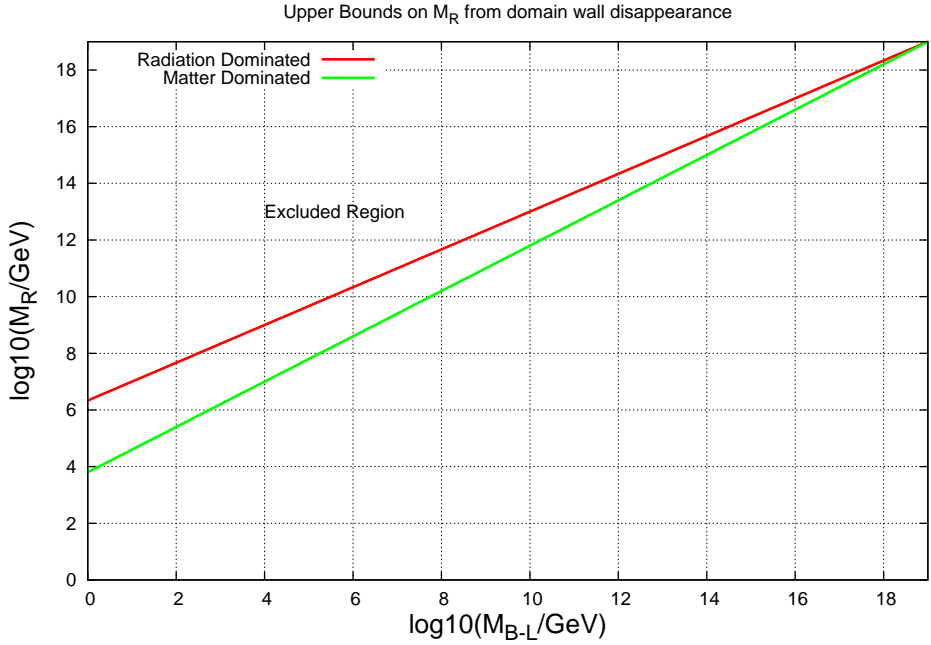}
\caption{Upper bound on $M_R$ from successful domain wall disappearance as a function of $M_{B-L}$}
\label{fig4}
\end{figure}
\subsection{Extension by a pair of Higgs Triplets: Introducing multi-step symmetry breaking}
In this section, we discuss another possibility to make the domain walls disappear in minimal LRSM by introducing 
an additional pair of Higgs triplets $\Omega_L (1,3,1,0), \Omega_R (1,1,3,0)$. As we will see, this extra pair of fields not only
provides a viable mechanism for domain wall disappearance, but also allows the possibility to achieve the symmetry breaking $SU(2)_R \times U(1)_{B-L} \rightarrow U(1)_Y$ at two different stages:
$$ SU(2)_L \times SU(2)_R \times U(1)_{B-L} \quad \underrightarrow{\langle
\Omega_R \rangle} \quad SU(2)_L \times U(1)_R \times U(1)_{B-L} \quad \underrightarrow{\langle \Sigma_R \rangle} \quad SU(2)_L\times U(1)_Y $$ 
where $\Sigma$ can either be a Higgs doublet or a Higgs triplet. Unlike in the case of minimal LRSM, here we can have dimension five Planck suppressed terms in the scalar potential which, as we will see, create sufficient pressure difference across the domain walls to make them disappear. These operators can be written as
\begin{equation}
V_{\text{NR}} \supset \frac{f_L}{M_{Pl}} (\Omega^{\dagger}_L \Omega_L )(\Sigma^{\dagger}_L \Omega_L \Sigma_L ) + \frac{f_R}{M_{Pl}} (\Omega^{\dagger}_R \Omega_R )(\Sigma^{\dagger}_R \Omega_R \Sigma_R )
\end{equation} 
We denote the vev of $\Sigma_R$ as the scale of $U(1)_{B-L}$ symmetry breaking $M_{B-L}$ and that of $\Omega_R$ as $M_R$. Assuming non-zero vev for only the right handed Higgs fields and zero vev for the left handed ones, the effective potential becomes
\begin{equation}
V^R_{eff}  \sim \frac{f_R}{M_{Pl}}M^2_{B-L} M^3_R 
\end{equation}
Similarly, assuming only the left type Higgs to acquire non-zero vev, we have the effective potential as
\begin{equation}
V^L_{eff}  \sim \frac{f_L}{M_{Pl}}M^2_{B-L} M^3_R 
\end{equation}
The energy difference across the walls separating the left and the right sectors is nothing but $V^L_{eff}-V^R_{eff}$:
\begin{equation}
\delta \rho \sim \frac{(f_L-f_R)}{M_{Pl}}M^2_{B-L} M^3_R 
\end{equation}
Comparing this $\delta \rho$ with the pressure difference needed for the domain walls to disappear during a matter dominated era, we get
\begin{equation}
 \frac{(f_L-f_R)}{M_{Pl}}M^2_{B-L} M^3_R > M_R^4 \left(\frac{M_R}{M_{Pl}}\right)^{3/2}
\end{equation}
Taking the dimensionless parameters $f$ to be of order unity, the above equation gives a upper bound
on $M_R$ in a matter dominated era
\begin{equation}
M_R < (M_{Pl}M^4_{B-L})^{1/5}
\end{equation}
Similarly for radiation dominated era
\begin{equation}
\frac{(f_L-f_R)}{M_{Pl}}M^2_{B-L} M^3_R  >M_R^4 \left(\frac{M_R}{M_{Pl}}\right)^2
\end{equation}
which gives a similar upper bound on $M_R$ as
\begin{equation}
M_R < (M_{Pl}M^2_{B-L})^{1/3}
\end{equation}
Thus, for a TeV scale $M_{B-L}$, the scale of parity breaking $M_R$ has to be less than $10^6 \; \text{GeV}$ and $10^8 \; \text{GeV}$ for matter and radiation dominated era respectively. Comparing the obtained $\delta \rho$ with the weak inflation case we have 
\begin{equation}
\frac{(f_L-f_R)}{M_{Pl}}M^2_{B-L} M^3_R  \geq  M^4_R \frac{T^{12}_D M^3_{Pl}}{M^{15}_R}
\end{equation}
Taking the dimensionless coefficients to be of order one, we arrive at the following bound on $M_R$
\begin{equation}
M_R \geq  2.7 \times 10^{5} T^{2/7}_D M^{-1/7}_{B-L} 
\end{equation}

\section{Gauge Coupling Unification}
\label{GUT}
The one-loop renormalization group evolution equations \cite{Jones:1981we} are given by 
\begin{equation}
\mu \frac{ d g_i}{d\mu} = \beta_i(g_i) = \frac{g^3_i}{16 \pi^2} b_i
\label{eq6}
\end{equation}
Defining $\alpha_i = g^2_i/(4\pi) $ and $ t = ln( \mu/\mu_0) $ and the most general renormalization group equation above becomes
\begin{equation}
\frac{d \alpha^{-1}_i}{dt} = -\frac{b_i}{2 \pi}
\label{eq7}
\end{equation}
The one-loop beta function is given by
\begin{equation}
\beta_i(g_i)=\frac{g^3_i}{16 \pi^2}[-\frac{11}{3}\text{Tr}[T^2_a]+\frac{2}{3}\sum_{f} \text{Tr}[T^2_f]+\frac{1}{3}\sum_{s} \text{Tr}[T^2_s]]
\end{equation}
where $f$ and $s$ denote the fermions and scalars respectively. For $SU(N)$, Tr$[T^2_a] =N$ and Tr$[T_i T_i]=\frac{1}{2}$. We closely follow the analysis in \cite{Lindner} to calculate the beta functions in both the minimal as well as extended LRSM discussed above. The experimental initial values for the couplings at electroweak scale $M=M_Z$ \cite{Amsler:2008zzb} are
\begin{equation}
\left(\begin{array}{cc}
\ \alpha_3 (M_Z) \\
\ \alpha_2(M_Z) \\
\ \alpha_1(M_Z)
\end{array}\right)
= \left(\begin{array}{cc}
\ 0.118 \pm 0.003 \\
\ 0.033493^{+0.000042}_{-0.000038} \\
\ 0.016829 \pm 0.000017
\end{array}\right)
\label{par1}
\end{equation}
The normalization condition at $M = M_{R} \;(M = M_{B-L})$ where the $U(1)_Y$ gauge coupling
merge with $SU(2)_R \times U(1)_{B-L} \; (U(1)_R \times U(1)_{B-L}) $ is $\alpha^{-1}_{B-L}=\frac{5}{2}
\alpha^{-1}_Y-\frac{3}{2}\alpha^{-1}_L$. Using all these, the gauge coupling unification for minimal LRSM with doublets and triplets are shown in figure \ref{fig1}, \ref{fig2} respectively. We show that, just with the minimal field content and no additional fields, successful gauge coupling unification can be achieved only when the parity breaking scale $M_R$ is as high as $3 \times 10^{11} \; \text{GeV}$ and $ 3 \times 10^9 \; \text{GeV}$ for doublet and triplet model respectively. However, if extra pairs of fields $\chi (1,1,1,2), \bar{\chi}(1,1,1,-2)$ are taken into account, the scale of parity breaking $M_R$ can be as low as a few TeV's as can be seen from figure \ref{fig1}, \ref{fig2}. These extra fields although looks unnatural (if large in numbers), can naturally fit inside $SO(10)$ representations like $\bf{120}$.

Similarly, for the extended LRSM discussed in section \ref{extLR} we study the gauge coupling unification and show that with just $\Delta_{L,R}, \Omega_{L,R}$ as the Higgs content apart from the usual bidoublet, gauge coupling unification can be achieved for the symmetry breaking scales $M_{B-L} = 3\; \text{TeV}, M_R = 10^8 \; \text{GeV}$ as can be seen from figure \ref{fig3}. Such a possibility of a TeV scale $U(1)_{B-L}$ gauge symmetry is quite tantalizing in view of the ongoing collider experiments like Large Hadron Collider (LHC). The scale of parity breaking $M_R$ can be lowered further by incorporating additional fields like $\chi, \bar{\chi}$ as discussed in the case of minimal LRSM. We find that for three such additional pairs of fields, both $M_{B-L}$ and $M_R$ can be as low as a few TeV's.
\begin{center}
\begin{table}[ht]
\caption{Bounds on $M_R/\text{GeV}$ in Left Right Symmetric Models}
\begin{tabular}{|c|c|c|c|c|}
\hline
Model  & Gauge Coupling  & DW removal & DW removal & DW removal \\
      &   Unification      & during MD era    & during RD era  & including WI\\
\hline
Minimal & $ \sim 3\times 10^{11} $ & $  > M_{Pl}$ & $f_L-f_R>1$ & $\geq  3.7 \times 10^{13} T^{12/17}_D$ \\
Doublet & & & & \\
\hline
Minimal & $ \sim 3\times 10^{9} $ & $  > M_{Pl}$ & $f_L-f_R>1$ & $\geq 3.7 \times 10^{13} T^{12/17}_D $ \\
Triplet & & & & \\
\hline
Extended & $\sim 3\times (10^9-10^{11})$ & $< 10^{17} $ & $< 10^{17} $ & $\geq 1 \times 10^{4} T^{4/5}_D$ \\
Singlet & & & & \\
\hline
Extended & $ \sim 1 \times 10^8$ & $< (M_{Pl}M^4_{B-L})^{1/5}$ & $< (M_{Pl}M^2_{B-L})^{1/3}$ &  $\geq  2.7 \times 10^{5} T^{2/7}_D M^{-1/7}_{B-L}$ \\
Triplet & & & & \\
\hline
\end{tabular}
\label{table1}
\end{table}
\end{center} 

\section{Results and Conclusion}
\label{conclude}
We have studied the domain wall formation as a result of spontaneous breaking of a discrete symmetry called D parity in generic left right symmetric models. Since stable domain walls 
are in conflict with cosmology, we consider the effects of Planck scale suppressed operators 
in destabilizing them. We consider the evolution and decay of domain walls in two different
epochs: radiation dominated as well as matter dominated. We find that in minimal versions of these models, the successful removal of 
domain walls put such constraints on the D-parity breaking scale $M_R$, which are not possible to realize in any physical theory, for example $M_R > M_{Planck}$. We also study gauge coupling unification in minimal versions of these models and find that with the minimal field content $M_R$ has to be as high as $ 10^9-10^{11} \; \text{GeV}$ (far beyond the reach of present experiments) for successful gauge coupling unification to be achieved. 

To have successful domain wall disappearance as well as to explore the possibility of a TeV scale intermediate symmetry, we study two viable extension of minimal LRSM : one with a gauge singlet and one with a pair of triplets with $U(1)_{B-L}$ charge zero. A gauge singlet although do not affect the running of gauge coupling, contributes to the effective energy density in such a way that sufficient pressure difference can be created across the domain walls to make them disappear without having some unphysical constraints like $M_R>M_{Planck}$ as in the case of minimal LRSM. Extension by Higgs triplets $\Omega_{L,R}$ not only provides a solution to the domain wall problem, but also allows the possibility to have separate $SU(2)_R$ and $U(1)_{B-L}$ symmetry breaking scales. This allows us to have a TeV scale $U(1)_{B-L}$ symmetry even if the scale of parity breaking $M_R$ is restricted to be as high as $10^8 \; \text{GeV}$. However, to agree with the constraints coming from domain wall disappearance ($M_R < 10^6-10^8 \; \text{GeV}$ for TeV scale $M_{B-L}$), we have to incorporate additional pairs of fields $\chi (1,1,1,2), \bar{\chi}(1,1,1,-2)$ as seen from figure \ref{fig3}. These extra pairs of fields can naturally fit inside $SO(10)$ representations like $\bf{120}$. The constraints on the scale of parity breaking $M_R$ from domain wall disappearance in this model also depends on the scale of $U(1)_{B-L}$ symmetry breaking $M_{B-L}$. We show the upper bound on $M_R$ as a function of $M_{B-L}$ in figure \ref{fig4} for both matter and radiation dominated era. A brief summary of our results is presented in the table \ref{table1}.

To summarize the results of our paper, we have shown that in the minimal versions of non-supersymmetric left right models, successful domain wall disappearance can not be achieved. However, we can achieve successful gauge coupling unification with the minimal field content if the scale of parity breaking is as high as $10^9-10^{11}\; \text{GeV}$ which can be lowered further by incorporating additional fields. Hence suitable extensions of minimal LRSM is required to achieve both domain wall disappearance as well as gauge coupling unification together with the tantalizing possibility of a TeV scale intermediate symmetry. We propose two such extensions and show that in Higgs triplet extension of minimal LRSM, all these possibilities can be realized simultaneously.

\section{Acknowledgment}
We would like to thank Prof. Urjit A. Yajnik, IIT Bombay for useful comments and discussions.

\end{document}